\documentclass[12pt]{article}
\begin{document}
\centerline{{\bf ON A REPRESENTATION OF
MATHISSON-PAPAPETROU-DIXON}} \centerline{{\bf EQUATIONS IN THE
KERR METRIC}}

\vspace{3mm}

\centerline{R M  Plyatsko, O B Stefanyshyn  and M T Fenyk}

\vspace{2mm}

\centerline{ Pidstryhach Institute for Applied Problems in
Mechanics and Mathematics} \centerline{Ukrainian National Academy
of Sciences, 3-b Naukova Str.,} \centerline{Lviv, 79060, Ukraine}

\vspace{3mm}

\centerline{{\bf Abstract}}

\vspace{2mm}

New representation of the exact Mathisson-Papapetrou-Dixon
equations at the Mathisson-Pirani condition in the Kerr metric
which does not contain the third-order derivatives of the
coordinates of a spinning particle is obtained. For this purpose
the integrals of energy and angular momentum of the spinning
particle as well as a differential relationship following from the
Mathisson-Papapetrou-Dixon equations are used. The form of these
equations is adapted for their computer integration with the aim
of further investigations of the influence of the spin-curvature
interaction on the particle's behavior in the gravitational field
without restrictions on its velocity and spin orientation.

\vspace{2mm}


\vspace{5mm}

\centerline{{\bf 1. Introduction}}

\vspace{3mm}

In general relativity the two main approaches have been developed
for the description of the spinning particle behavior in the
gravitational field. Chronologically the first of them was
initiated in 1929 when the usual Dirac equation was generalized
for the curved space-time [1]. The second, the pure classical
(non-quantum) approach, has been proposed in 1937 [2]. Later it
was shown that in the certain sense the equations from [2] follow
from the general relativistic Dirac equation as some classical
approximation [3].

In the focus of this paper are the equations of motion of the
classical spinning particle, which after [2] were obtained in [4]
and later in many papers by different methods. These equations can
be written as
\begin{equation}\label{1}
\frac D {ds} \left(mu^\lambda + u_\mu\frac {DS^{\lambda\mu}}
{ds}\right)= -\frac {1} {2} u^\pi S^{\rho\sigma}
R^{\lambda}_{~\pi\rho\sigma},
\end{equation}
\begin{equation}\label{2}
\frac {DS^{\mu\nu}} {ds} + u^\mu u_\sigma \frac {DS^{\nu\sigma}}
{ds} - u^\nu u_\sigma \frac {DS^{\mu\sigma}} {ds} = 0,
\end{equation}
where $u^\lambda\equiv dx^\lambda/ds$ is the particle's
4-velocity, $S^{\mu\nu}$ is the tensor of spin, $m$ and $D/ds$
are, respectively, the mass and the covariant derivative with
respect to the particle's proper time $s$;
$R^{\lambda}_{~\pi\rho\sigma}$ is the Riemann curvature tensor
(units $c=G=1$ are used); here and in the following, Latin indices
run 1, 2, 3 and Greek indices 1, 2, 3, 4; the signature of the
metric (--,--,--,+) is chosen.

Equations (1), (2) where generalized in [5] for the higher
multipoles of the test particles and now set (1), (2) is known as
the Mathisson-Papapetrou-Dixon (MPD) equations.

The first effects of the spin-gravity interaction following from
(1)--(2) have been considered in [6] for the Schwarzschild field.
According to [6] and by many further publications (some list of
them is presented, for example, in [7, 8]) the influence of spin
on the particle's trajectory is negligible small for practical
registrations. However, in this sense much more realistic are the
effects connected with the spin precession [9].

The interesting point has been elucidated in [10] concerning the
possibility of the static position of a spinning particle outside
the equatorial plane of the Kerr source of the gravitational
field, on its axis of rotation. In spite of the conclusion that
such situation is not allowed by the MPD equations [10], this
question stimulated the investigations of possibilities of some
non-static (dynamical) effects connected with the particle's
motion relative to a Schwarzschild or Kerr mass outside the
equatorial plane [11]. Then it was shown that spinning particles
moving with relativistic velocity can deviate from geodesics
significantly [12, 13].

While investigating the solutions of equations (1), (2), it is
necessary to add a supplementary condition  in order to choose an
appropriate trajectory of the particle's center of mass. Most
often the conditions [2, 14]
\begin{equation}\label{3}
S^{\lambda\nu} u_\nu = 0
\end{equation}
or [5, 15]
\begin{equation}\label{4}
S^{\lambda\nu} P_\nu = 0
\end{equation}
are used, where
\begin{equation}\label{5}
P^\nu = mu^\nu + u_\lambda\frac {DS^{\nu\lambda}}{ds}
\end{equation}
is the 4-momentum. The condition for a spinning test particle
\begin{equation}\label{6} \frac{|S_0|}{mr}\equiv\varepsilon\ll 1
\end{equation}
must be taken into account as well [10], where  $|S_0|=const$ is
the absolute value of spin, $r$ is the characteristic length scale
of the background space-time (in particular, for the Kerr metric
$r$ is the radial coordinate), and $S_0$ is determined by the
relationship
\begin{equation}\label{7} S_0^2=\frac12
S_{\mu\nu}S^{\mu\nu}.
\end{equation}

In general, the solutions of equations (1), (2) under conditions
(3) and (4) are different. However, in the post-Newtonian
approximation these solutions coincide with high accuracy, just as
in some other cases [16, 17]. Therefore, instead of exact MPD
equations (1) their linear spin approximation
\begin{equation}\label{8}
m\frac D {ds} u^\lambda = -\frac {1} {2} u^\pi S^{\rho\sigma}
R^{\lambda}_{\,\,\,\pi\rho\sigma}
\end{equation}
is often considered. In this approximation condition (4) coincides
with (3) (by condition (3)in equations (1) $m=const$).

According to [18] for a massless spinning particle, which moves
with the velocity of light, the appropriate condition is (3).
Which condition is adequate for the motions of a spinning particle
with the nonzero mass if its velocity is close to the velocity of
light? To answer this question it is necessary to analyze the
corresponding solutions of the exact MPD equations (1), (2) both
at condition (3) and (4).

The main purpose of this paper is to consider the exact MPD
equations under condition (3) in a Kerr metric. We use this metric
in the Boyer-Lindquist coordinates $x^1=r, \quad x^2=\theta, \quad
x^3=\varphi, \quad x^4=t.$ There are Killing vectors $\vec \xi$
due to the symmetry of the Kerr space-time. As a result, equations
(1), (2) have the constants of motion $C_{\xi}$:
$$
C_{\xi}=\xi^\mu P_\mu - \frac12 \xi_{\mu; \nu} S^{\mu\nu},
$$
and from it follows [19, 28--31]
\begin{equation}\label{9}
E=P_4-\frac{1}{2}g_{4\mu,\nu}S^{\mu\nu},
\end{equation}
 \begin{equation}\label{10}
J_z=-P_3+\frac{1}{2}g_{3\mu,\nu}S^{\mu\nu},
\end{equation}
where $E$ and $J_z$ are the constants of the particle's energy and
the projection of the angular momentum correspondingly.

If $S^{\mu\nu}=0$, expressions (9), (10) pass to the known
relationships for the geodesic motions which were effectively used
for analyzing possible orbits of a spinless particle in a Kerr
space-time [20, 21]. Namely, by the constants of energy and
angular momentum the standard form of the geodesic equations,
which are the differential equations of the second-order by the
coordinates, can be reduced to the differential equations of the
first order. Naturally, it is interesting to apply the similar
procedure to the exact MPD equations using equations (9), (10).
However, in contrast to the geodesic equations, the exact MPD
equations at the condition (3) contain the third derivatives of
the coordinates [22, 23]. Therefore, the application of this
procedure to the exact MPD equations is not trivial.

In this paper for obtaining full set of the MPD equations at
condition (3) without the third derivatives of the coordinates
some differential relationship following from equations (1), (2)
are used. We present this relationship in section 2 in general
form, for any metric. Its concrete form in the Kerr metric,
together with equations (9), (10), is used in section 3 and the
full set of the differential equations for the dimensionless
quantities connected with the particle's coordinates, velocity and
spin is described. We analyze the relationship between $u^\lambda$
and $P^\lambda$ at condition (4) in section 4.  We conclude in
section 5.

It is known [22, 23] that in the Minkowski space-time the exact
MPD equations under condition (3) have, in addition to usual
solutions describing the straight worldlines, a set of solutions
describing the oscillatory (helical) worldlines. The physical
interpretation of these superfluous solutions was proposed by C.
M{\"o}ller [24]. He pointed out that in relativity the position of
the center of mass of a rotating body depends on the frame of
reference, and condition (3) is common for the so-called proper
and non-proper centers of mass [24]. The usual solution describe
the motion of the proper center of mass and the helical solutions
describe the motions of the set of the non-proper centers of mass.
Naturally, in general relativity, when the gravitational field is
present, the exact MPD equations (1)--(3) have some superfluous
solutions as well. Just to avoid these solutions, instead of (3)
condition (4) was used in many papers. In contrast to (3),
condition (4) picks out the unique worldline of a spinning
particle in the gravitational field. However, the question arises:
is this worldline close, in the certain sense, to the usual
(non-helical) worldline of equations (1), (2) under condition (3)?
It is simple to answer this question if the linear spin
approximation is valid, because in this case condition (4)
practically coincides with (3). Whereas another situation cannot
be excluded {\it a priori} for the high particle's velocity.

Note that the very condition (3) arose in a natural fashion in the
course of its derivation by different methods [25--27]. Therefore,
it is of importance to obtain a representation of the exact MPD
equations at this condition in the Kerr metric convenient for
their further computer integration.

We point out that the integrals of energy and angular momentum of
the MPD equations in a Kerr space-time were effectively used for
different purposes in [7, 28--36] at condition (4).

\vskip 5mm

\centerline{{\bf 2. A relationship following from equations
(1)--(3)}}

\vspace{3mm}

In addition to the antisymmetric tensor $S^{\mu\nu}$ in many
papers the 4-vector of spin $s_\lambda$ is used as well, where by
definition
\begin{equation}\label{11}
s_\lambda=\frac12
\sqrt{-g}\varepsilon_{\lambda\mu\nu\sigma}S^{\nu\sigma}
\end{equation}
and $g$ is the determinant of the metric tensor,
$\varepsilon_{\lambda\mu\nu\sigma}$ is the Levi-Civita symbol. It
follows from (7), (11) that $s_\lambda s^\lambda=S_0^2$ and at
condition (3) we have $s_\lambda u^\lambda=0$ (other useful
relationships with $s_\lambda$ following from MPD equations at
different supplementary condition can be found, for example, in
[8]).

The set of equations (2) contains three independent differential
equations and in (3) we have three independent algebraic
relationships between $S^{\lambda\nu}$ and $u_\mu$. By (3) the
components $S^{i4}$ can be expressed through $S^{ki}$:
\begin{equation}\label{12}
S^{i4}=\frac{u_k}{u_4}S^{ki}.
\end{equation}
So, using (12) the components $S^{i4}$ can be eliminated both from
equations (2) and (1). That is, in further consideration one can
"forgot" about supplementary condition (3) and deal with the three
independent components $S^{ik}$. However, it is appear that more
convenient form of equations (1), (2) is not for $S^{ik}$ but for
another 3-component value  $S_i$ which is connected with $S^{ik}$
by the simple relationship
\begin{equation}\label{13} S_i=\frac{1}{2u_4}
\sqrt{-g}\varepsilon_{ikl}S^{kl},
\end{equation}
 where
$\varepsilon_{ikl}$ is the spatial Levi-Civita symbol. For
example, it is not difficult to check that three independent
equations of set (2) in terms of $S_i$ can be written as
\begin{equation}\label{14}
u_{4} \dot S_i  + 2(\dot u_{[4} u_{i]} - u^\pi u_\rho
\Gamma^\rho_{\pi[4} u_{i]})S_k u^k + 2S_n \Gamma^n _{\pi [4}
u_{i]} u^\pi =0.
\end{equation}

The simple calculation shows that the 3-component value $S_i$ has
the 3-vector properties relative to the coordinate transformations
of the partial form $\hat x^i=\hat x^i(x^1, x^2, x^3), \quad \hat
x^4=x^4$ and in this special sense $S_i$ can be called as a
3-vector. (By the way, in the context of equations (1), (2)
firstly the 3-vector of spin was used in [6] with the notation
${\bf S}=(S^{23}, S^{31}, S^{12})$). By equations (11)--(13) the
relationship between $S_i$ and $s_\lambda$ is
\begin{equation}\label{15} S_i=-s_i+\frac{u_i}{u_4}s_4.
\end{equation}

Let us consider the first three equations of the subset (1) with
the indexes $\lambda=1, 2, 3$. Multiplying these equations by
$S_1,\quad S_2,\quad S_3$ correspondingly and taking into account
(11)--(13) we get
\begin{equation}\label{16}
mS_i\frac{Du^i}{ds}= -\frac12 u^\pi
S^{\rho\sigma}S_jR^j_{~\pi\rho\sigma}.
\end{equation}
We stress that in contrast to the each equation from set (1),
which contain the third derivatives of the coordinates,
relationship (16) does not have these derivatives.

Relationship (16) is an analog of relationship (21) from [8] where
the spin 4-vector $s_\lambda$ is used.

\vspace{5mm}

\centerline{{\bf 3. On set of exact MPD equations}}
\centerline{{\bf with constants
 of motion $E$, $J_z$ for the Kerr metric }}

\vspace{3mm}

In the Boyer-Lindquist coordinates the non-zero
 components of the Kerr metric tensor are
 \[
g_{11}=-\frac{\rho^2}{\Delta}, \quad g_{22}=-\rho^2,
\]
\[
g_{33}=-\left(r^2+a^2+\frac{2Mra^2}{\rho^2}
\sin^2\theta\right)\sin^2\theta,
\]
\begin{equation}\label{17}
 g_{34}=\frac{2Mra}{\rho^2}\sin^2\theta, \quad
g_{44}=1-\frac{2Mr}{\rho^2},
\end{equation}
where
\[
 \rho^2=r^2+a^2\cos^2\theta, \quad \Delta=r^2-2Mr+a^2, \quad
0\le\theta\le\pi.
\]

It is convenient to use the dimensionless quantities $y_i$
connected with the  particle's coordinates, where by definition
\begin{equation}\label{18}
\quad y_1=\frac{r}{M},\quad y_2=\theta,\quad y_3=\varphi, \quad
y_4=\frac{t}{M},
\end{equation}
as well as the quantities connected with its 4-velocity
\begin{equation}\label{19}
y_5=u^1,\quad y_6=Mu^2,\quad y_7=Mu^3,\quad y_8=u^4
\end{equation}
and the spin components [12]
\begin{equation}\label{20}
    y_9=\frac{S_1}{mM},\quad y_{10}=\frac{S_2}{mM^2},\quad
    y_{11}=\frac{S_3}{mM^2}.
\end{equation}
In addition, we introduce the dimensionless quantities connected
with the particle's proper time $s$ and the constants of motion
$E$, $J_z$ which is presented in (9), (10):
\begin{equation}\label{21}
    x=\frac{s}{M}, \quad \hat E=\frac{E}{m},\quad
    \hat J=\frac{J_z}{mM}.
\end{equation}
Quantities (18), (19) satisfy the four simple equations
\begin{equation}\label{22}
\dot y_1=y_5, \quad \dot y_2=y_6, \quad \dot y_3=y_7, \quad \dot
y_4=y_8,
\end{equation}
here a dot denotes the usual derivative with respect to $x$.

Now we point out the seven nontrivial first-order differential
equations for the 11 functions $y_i$. Namely, the first of them
follows directly from equation (16). The second is a result of the
covariant differentiation of the normalization condition $u_\nu
u^\nu=1$, that is
\begin{equation}\label{23}
u_\nu\frac{Du^\nu}{ds}=0.
\end{equation}
The third and fourth equations follow from (9) and (10)
correspondingly. Finally, the last three equations for $y_i$
follow from (2) as written for the 3-vector of spin (we recall
that for any metric the set of  equations (2) contains three
independent differential equations). Naturally, this set of the
seven equations is too long and is not presented here for brevity.
These seven equations together with the four equations from (22)
are the full set of the exact MPD equations which describe most
general motions of a spinning particle in the Kerr gravitational
field without any restrictions on its velocity and spin
orientation. We stress that the two equations following from (9)
and (10) contain the quantities $\hat E$ and $\hat J$ as the
parameters proportional to the particle's energy and angular
momentum according to notation (21). By choosing different values
of $\hat E$ and $\hat J$ for the fixed initial values of $y_i$ one
can describe the motions of different centers of mass. Among the
set of the pairs $\hat E$ and $\hat J$ there is the single pair
corresponding to the proper center of mass. The possible
approaches for finding this pair is a separate subject. One of
them was proposed in [37] where a method of separation of some
nonoscillatory solutions of the exact MPD equations in the
Schwarzschild field was considered. In the next section we shall
analyze the possibility of using condition (4) for the same
purpose.

\vspace{5mm}

\centerline{{\bf 4. Values $\hat E$ and $\hat J$ according to
condition (4)}}

\vspace{3mm}

Let us check the supposition that the single solutions of
equations (1), (2) at the supplementary condition (4),
corresponding to the fixed initial values of the coordinates,
velocity and spin, is close to those solutions of equations (1),
(2) at the condition (3) which describe the motion of the proper
center of mass with the same initial values. As we pointed out in
section 1, this assumption is justified for the velocities which
are not close to the velocity of light. Here we shall consider the
situation for any velocity.

Let us write the main relationships following from the MPD
equations at condition (4) [28--31]. The mass of a spinning
particle $\mu$ is defined as
\begin{equation}\label{24}
\mu=\sqrt{P_\lambda P^\lambda}
\end{equation}
and $\mu$ is the integral of motion, that is $d\mu/ds=0$. The
quantity $V^\lambda$ is the normalized momentum, where by
definition
\begin{equation}\label{25}
V^\lambda=\frac{P^\lambda}{\mu}.
\end{equation}
Sometimes $V^\lambda$ is called the "dynamical 4-velocity",
whereas the quantity $u^\lambda$ from (1)--(3) is the "kinematical
4-velocity" [7]. As the normalized quantities $u^\lambda$ and
$V^\lambda$ satisfy the relationships
\begin{equation}\label{26}
u_\lambda u^\lambda=1,\quad V_\lambda V^\lambda=1.
\end{equation}
There is the important relationship between $u^\lambda$ and
$V^\lambda$ [28--30]:
\begin{equation}\label{27}
    u^{\lambda}=N\left[V^\lambda+\frac{1}{2\mu^2\Delta}
    S^{\lambda\nu}V^{\pi}R_{\nu\pi\rho\sigma}S^{\rho\sigma}\right],
\end{equation}
where
\begin{equation}\label{28}
\Delta=1+\frac{1}{4\mu^2}R_{\lambda\pi\rho\sigma}S^{\lambda\pi}S^{\rho\sigma},
\end{equation}

Now our aim is to consider the explicit form of expression (27)
for the concrete case of the Schwarzschild metric, for the
particle motion in the plane $\theta=\pi/2$ when spin is
orthogonal to this plane (we use the standard Schwarzschild
coordinates $x^1=r, \quad x^2=\theta, \quad x^3=\varphi, \quad
x^4=t$). Then we have
\begin{equation}\label{29}
u^2=0,\quad u^1\ne 0, \quad u^3\ne 0, \quad u^4\ne 0,
\end{equation}
\begin{equation}\label{30}
S^{12}=0,\quad S^{23}=0,\quad S^{13}\ne 0.
\end{equation}
In addition to (30) by condition (4) we write
\begin{equation}\label{31}
 S^{14}=-\frac{P_3}{P_4}S^{13},\quad S^{24}=0, \quad S^{34}=\frac{P_1}{P_4}S^{13}.
\end{equation}
Using (7), (29)--(31) and the corresponding expressions for the
Riemann tensor in the Schwarzschild metric, from (27) we obtain
\[
u^1=NV^1\left(1+\frac{3M}{r^3}V_3
V^3\frac{S_0^2}{\mu^2\Delta}\right), \quad u^2=V^2=0,
\]
\[
u^3=NV^3\left[1+\frac{3M}{r^3}(V_3
V^3-1)\frac{S_0^2}{\mu^2\Delta}\right],
\]
\begin{equation}\label{32}
u^4=NV^4\left(1+\frac{3M}{r^3}V_3
V^3\frac{S_0^2}{\mu^2\Delta}\right),
\end{equation}
where $M$ is the mass of a Schwarzchild source. According to (28)
we write the expression for $\Delta$ as
\begin{equation}\label{33}
\Delta=1+\frac{S_0^2 M}{\mu^2 r^3}(1-3V_3 V^3)
\end{equation}
(the quantity $M$ in (32), (33) is the mass of a Schwarzchild
source). Inserting (33) into (32) we get
\[
u^1=\frac{NV^1}{\Delta}\left(1+\frac{S_0^2 M}{\mu^2 r^3}\right),
\]
\[u^3=\frac{NV^3}{\Delta}\left(1-2\frac{S_0^2 M}{\mu^2 r^3}\right),
\]
\begin{equation}\label{34}
u^4=\frac{NV^4}{\Delta}\left(1+\frac{S_0^2 M}{\mu^2 r^3}\right).
\end{equation}
In the following we use the notation
\begin{equation}\label{35}
\varepsilon=\frac{|S_0|}{\mu r},
\end{equation}
where according to the condition for a test particle it is
necessary $\varepsilon\ll 1.$ However, in our calculations we
shall keep all terms with $\varepsilon.$

 The explicit expressions for $N$ we obtain directly from the condition
$u_\lambda u^\lambda=1$ in the form
\[
N=\Delta\left[\left(1+\varepsilon^2\frac{M}{r}\right)^2-3V_3
V^3\varepsilon^2\frac{M}{r}\times\right.
\]
\begin{equation}\label{36}
\left.\times\left(2-\varepsilon^2\frac{M}{r}\right)\right]^{-1/2}.
\end{equation}
Inserting (36) into (34) we obtain the expression for the
components $V^\lambda$ through $u^\lambda$ ($V^2=u^2=0$):
\[
V^1=u^1R\left(1-2\varepsilon^2\frac{M}{r}\right),
\]
\[
V^3=u^3R\left(1+\varepsilon^2\frac{M}{r}\right),
\]
\begin{equation}\label{37}
V^4=u^4R\left(1-2\varepsilon^2\frac{M}{r}\right),
\end{equation}
where
\begin{equation}\label{38}
R=\left[\left(1-2\varepsilon^2\frac{M}{r}\right)^2-3(u^3)^2\varepsilon^2
Mr\left(2-\varepsilon^2\frac{M}{r}\right)\right]^{-1/2}.
\end{equation}

The main feature of relationships (37), (38) is that for the high
tangential velocity of a spinning particle the values $V^1, V^3,
V^4$ become imaginary. Indeed, if
\begin{equation}\label{39}
|u^3|>\frac{1}{\varepsilon\sqrt{6Mr}},
\end{equation}
in (38) we have the square root of the negative value. [As writing
(39) we neglect the small terms of order $\varepsilon^2$; all
equations in this section before (39) and after (40) are strict in
$\varepsilon.$] Using the notation for the particle's tangential
velocity $u_{tang}\equiv ru^3$ by (39) we write
\begin{equation}\label{40}
|u_{tang}|>\frac{\sqrt{r}}{\varepsilon\sqrt{6M}}.
\end{equation}
According to estimates similar to those which are presented in
[13] if $r$ is not much greater than $M$, the velocity value of
the right-hand side of equation (40) corresponds to the particle's
highly relativistic Lorentz $\gamma$-factor of order
$1/\varepsilon$.

Probably, this fact that according to (25), (37)--(40) the
expressions for the components of 4-momentum $P^\lambda$ become
imaginary is an evidence that condition (4) cannot be used for the
particle's velocity which is very close to the velocity of light.
However, this point needs some additional consideration. In any
case, relationships (37)--(40) are of importance for authors which
investigate solutions of the MPD equations at condition (4). We
stress that many papers were devoted to study the planar or
circular motions of spinning particles in the Schwarzschild or
Kerr space-time at different supplementary conditions [6--13,
28--35, 38]. Equations (37)--(40) elucidate the new specific
features which arise for the highly relativistic motions.

The question of momentum or velocity normalization for a spinning
particle was discussed in [28, 29]. It is pointed out in [28] that
there exist a critical distance of minimum approach of a spinning
particle to the Kerr source where its velocity becomes space-like.
The new result of this section as compare to [28, 29] consists in
the conclusion that according to (37)--(40) only tangential
component of velocity is important in this case, not the radial
one, although the orbit is not necessarily circular.

It is interesting to check the possibility of using the values $E$
and $J_z$ as calculated by (37) for computing spinning particle
motions by the equations which are described in the previous
section if the particle's tangential velocity is much less than
the critical value from the right-hand side of (40). At condition
(4) the constants $E$ and $J_z$ for the equatorial motions in the
Schwarzschild field can be written as
\begin{equation}\label{41}
E=P_4+\frac{1}{2}g_{44,1}S^{14}=\mu
V_4-\frac{1}{2}g_{44,1}\frac{V_3}{V_4}S^{13},
\end{equation}
 \begin{equation}\label{42}
J_z=-P_3-\frac{1}{2}g_{33,1}S^{13}=-\mu
V_3-\frac{1}{2}g_{33,1}S^{13}.
\end{equation}
Using the dimensionless quantities $y_i$ as defined in (18), (19),
relationship (37) and the simple expression for $S^{13}$ through
$S_0$ from (41), (42) we obtain
\begin{equation}\label{43}
\hat
E=R\left[y_8\left(1-\frac{2}{y_1}\right)\left(1-\frac{2\varepsilon^2}{y_1}\right)+
\varepsilon y_7\left(1+\frac{\varepsilon^2}{y_1}\right)\right],
\end{equation}
\begin{equation}\label{44}
\hat J=R\left[y_1^2 y_7 \left(1+\frac{\varepsilon^2}{y_1}\right) +
\varepsilon y_1 y_8
\left(1-\frac{2}{y_1}\right)\left(1-\frac{2\varepsilon^2}{y_1}\right)\right],
\end{equation}
where similarly to (21) we note $\hat E=E/\mu$, $\hat J=J_z/\mu
M$. Then for the fixed initial values of the quantities $y_1, y_7,
y_8$ in (43), (44) we have the concrete values of $\hat E$ and
$\hat J$ which can be used for numerical integration of the exact
MPD equations. Some examples of such integration we shall consider
in another paper.

\vspace{5mm}

\centerline{{\bf 5. Conclusions}}

\vspace{3mm}

In this paper we describe the representation of the exact MPD
equations at supplementary condition (3) for a Kerr metric which
can be obtained by using the constants of the particle's motion,
the energy and angular momentum, together with the differential
consequence of these equations (16). The way of obtaining the
corresponding set of the 11 first-order differential equations is
presented in section 3. The possibility using expressions (43),
(44) in these equations to describe motions of a spinning particle
in the Schwarzschild space-time we plan to consider in other
publications, as well as  to present results of complex
investigating the highly relativistic motions of a spinning
particle in the Kerr space-time according to the exact MPD
equations.

\vspace{3mm}


\end{document}